\documentclass{svproc}
\usepackage{url}
\usepackage{graphicx}
\usepackage{amsfonts}
\usepackage{amsmath}
\usepackage{amssymb}

\begin{document}
\mainmatter              
\title{(Anti)Fragility and Convex Responses in Medicine}
\titlerunning{Convex Responses in Medicine}  
%
\author{Nassim Nicholas Taleb }
\authorrunning{Nassim Nicholas Taleb} 
%
\tocauthor{Nassim Nicholas Taleb}
\institute{Tandon School of Engineering, New York University, NY 11201, USA\\
\email{NNT1@nyu.edu}}

\maketitle              

\begin{abstract}
This paper applies risk analysis to medical problems, through the properties of nonlinear responses (convex or concave). It shows 1)  necessary relations between the nonlinearity of dose-response and the statistical properties of the outcomes, particularly the effect of the variance (i.e., the expected frequency of the various results and other properties such as their average and variations); 2)  The description of "antifragility" as a mathematical property for local convex response and its generalization and the designation "fragility" as its opposite, locally concave; 3) necessary relations between dosage, severity of conditions, and iatrogenics. \\

Iatrogenics seen as the tail risk from a given intervention can be analyzed in a probabilistic decision-theoretic way, linking probability to nonlinearity of response. There is a necessary two-way mathematical relation between nonlinear response and the tail risk of a given intervention. \\

In short we propose a framework to integrate the necessary consequences of nonlinearities in evidence-based medicine and medical risk management.
\keywords{evidence based medicine, risk management, nonlinear responses}
\end{abstract}

\paragraph{Comment on the notations}: we use $x$ for the dose, $S(x)$ for the response function to $x$ when is sigmoidal (or was generated by an equation that is sigmoidal), and $f(x)$ when it is not necessarily so.

\section{ Background}\label{Background}
\begin{figure}
\includegraphics[width=.9\textwidth]{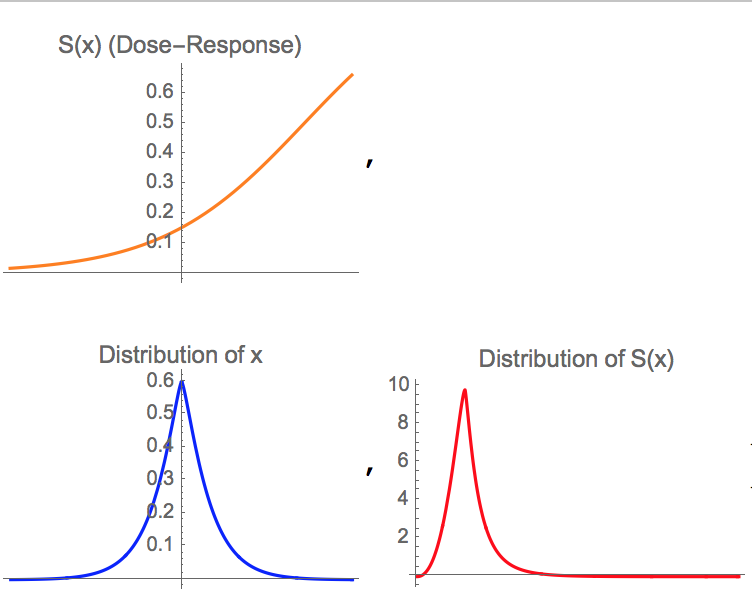}	
\caption{These two graphs summarize the gist of this chapter: how we can go from the reaction or dose-response $S(x)$, combined with the probability distribution of $x$, to the probability distribution of $S(x)$ and its properties: mean, expected benefits or harm, variance of $S(x)$. Thus we can play with the different parameters affecting $S(x)$ and those affecting the probability distribution of $x$, to assess results from output. $S(x)$ as we can see can take different shapes (We start with $S(x)$ monotone convex (top) or the second order sigmoid). }\label{DoseResponseDist}
\includegraphics[width=\textwidth]{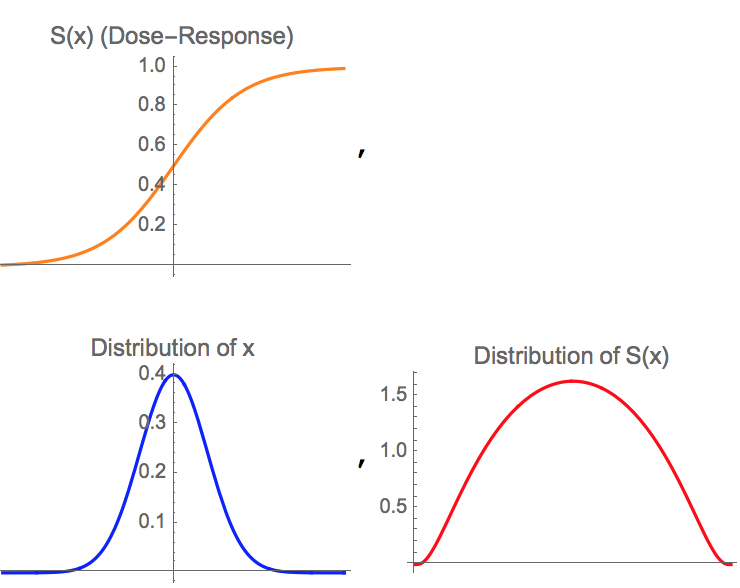}	
\end{figure}

Consideration of the probabilistic dimension has been made explicitly in some domains, for instance there are a few papers linking Jensen's inequality and noise in pulmonary ventilators: papers such as  Brewster et al. (2005)\cite{brewster2005convexity}, Graham et al.(2005) \cite{graham2005mathematical}, Funk (2004)\cite{funk2004comparison}, Arold et al. (2003)\cite{arold2003variable},  Mutch et al. (2007), Amato et al. \cite{amato1998effect}. In short, to synthesize the literature, continuous high pressures have been shown to be harmful with increased mortality, but occasional spikes of ventilation pressures can be advantageous with recruitment of collapsed alveoli, and do not cause further increased mortality. But explicit probabilistic formulations are missing in other domains, such as episodic energy deficit, intermittent fasting, variable  uneven distribution of sub-groups ( proteins and autophagy), vitamin absorption, high intensity training, fractional dosage, the comparative effects of chronic vs actute, moderate and distributed vs intense and concentrated, etc. %

 Further, the detection of convexity is still limited to local responses and does not appear to have led to decision-making under uncertainty and inferences on unseen risks based on the detection of nonlinearity in response, for example the relation between tumor size and the iatrogenics of intervention, or that between the numbers needed to treat and the side effects (visible and invisible) from an intervention such as statins or various blood pressure treatments.\\

 The links we are investigating are mathematical and necessary. And they are two-way (work in both directions). To use  a simple illustrative example:
\begin{itemize}
 	\item a convex response of humans to energy balance over a time window necessarily implies the benefits of intermittent fasting (seen as higher variance in the distribution of nutrients) over some range that time window,
 	\item the presence of misfitness in populations that have exceedingly steady nutrients, and evidence of human fitness to an environment that provides high variations (within bounds) in the availability of food, both necessarily imply a  nonlinear (concave) response to food over some range of intake and frequency (time window).
 \end{itemize}
 The point can be generalized in the same manner to energy deficits and the variance of the intensity of such deficits given a certain average.\\
 
 \textit{Note the gist of our approach: we are not asserting that the benefits of intermittent fasting or the existence of a convex response are true; we are just showing that if one is true then the other one is necessarily so, and building decision-making policies that bridge the two.}\\

Finally, note that convexity in medicine is at two levels. First, understanding the effect of dosing and its nonlinearity. Second, at the level of risk analysis for patients.

\subsection{Convexity and its Effects}
Let us define convexity as follows. Let the "response" function $f:\mathbb{R}^+ \rightarrow \mathbb{R} $ be a twice differentiable function. If over a range x $\in [a,b]$, over a set time period $\Delta t$, $\frac{\partial  ^2f(x)}{\partial  x^2}\geq $ 0, or more practically (by relaxing the assumptions of differentiability), $\frac{1}{2}(f(x+\Delta
x) + f(x-\Delta x))\geq f(x)$, with $x+\Delta x$ and $x-\Delta x \in [a,b]$ then there are benefits or harm from the unevenness of distribution, pending whether $f$ is defined as positive or favorable or modeled as a harm function (in which case one needs to reverse the sign for the interpretation).

In other words, in place of a dose $x$, one can give, say, 140\% of $x$, then 60\% of $x$, with a more favorable outcome one is in a zone that benefits from unevenness. Further, more unevenness is more beneficial: 140\% followed by 60\% produces better effects than, say, 120\% followed by 80\%.

We can generalize to comparing linear combinations:
$\sum\alpha_i =1$, $0\leq|\alpha_i|\leq 1$, $\sum(\alpha_i f(x_i)) \geq f(\sum(\alpha_i x_i))$; thus we end up with situations where, for $ x \leq b-\Delta$ and $n \in \mathbb{N}$, $f(n x) \geq n f(x)$. This last property describes a "stressor" as having higher intensity than zero: there may be no harm from $f(x)$ yet there will be one at higher levels of $x$.

\begin{figure}

	\includegraphics[width=\linewidth]{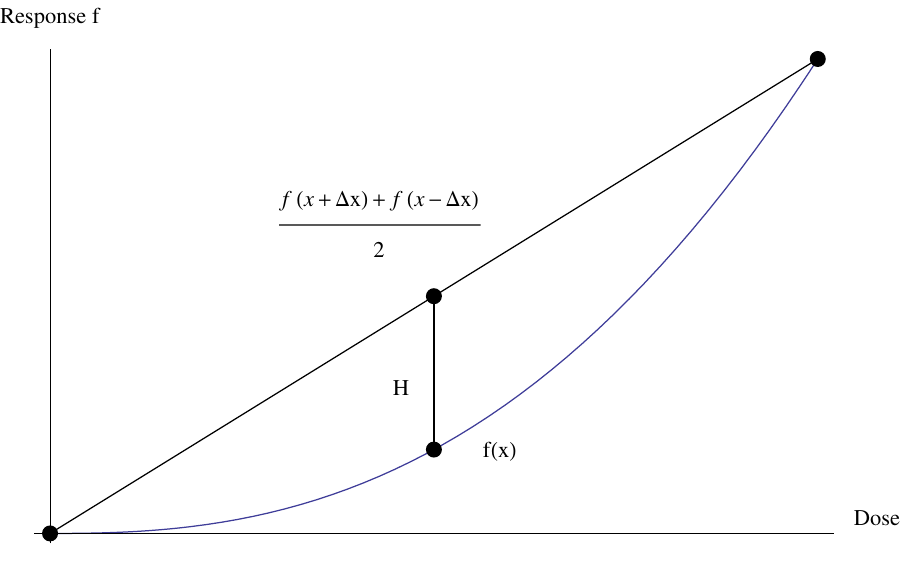}
	\caption{Jensen's inequality} \label{Jensen}
\end{figure}

Now if $X$ is a random variable with support in $[a,b]$ and $f$ is convex over the interval as per above, then 
\begin{equation}
 \mathbb{E}\left(f(x)\right) \geq f\left(\mathbb{E}(x)\right)\label{jensen}	, 
 \end{equation}
 what is commonly known as Jensen's Inequality, see Jensen(1906) \cite{jensen1906fonctions}, Fig. \ref{Jensen}. Further (without loss of generality), if its continuous distribution with density $\varphi(x)$ and support in $[a,b]$ belongs to the location scale family distribution, with $\varphi(\frac{x}{\sigma})= \sigma \varphi(x)$ and $\sigma>0$, then, with  $\mathbb{E}_{\sigma}$ the indexing representing the expectation under a probability distribution indexed by the scale $\sigma$, we have:
\begin{equation}
	\forall \sigma_2>\sigma_1,\, \mathbb{E}_{\sigma_2}\left(f(x)\right) \geq \mathbb{E}_{\sigma_1}\left(f(x)\right) \label{antifr}
\end{equation}
The last property implies that the convexity effect increases the expectation operator. We can verify that since $\int_{f(a)}^{f(b)} y \frac{\phi \left(f^{(-1)}(y)\right)}{f'\left(f^{(-1)}(y)\right)}\mathrm{d}y$
 is an increasing function of $\sigma$. A more simple approach (inspired from mathematical finance heuristics) is to consider for $0\leq\delta_1 \leq \delta_2\leq b-a$, where $\delta_1$ and $\delta_2$ are the mean expected deviations or, alternatively, the results of a simplified two-state system, each with probability $\frac{1}{2}$:
 \begin{equation}
 \frac{f(x-\delta_2)+	 f(x+\delta_2)}{2}\geq  \frac{f(x-\delta_1)+	 f(x+\delta_1)}{2} \geq f(x)
 \end{equation}

This is of course a simplification here since dose response is rarely monotone in its nonlinearity, as we will see in later sections. But we can at least make claims in a certain interval $[a,b]$.

\bigskip
 
\paragraph{What are we measuring?} Clearly, the dose (represented on the $x$ line) is hardly ambiguous: any quantity can do, such as pressure, caloric deficit, pounds per square inch, temperature, etc.

The response, harm or benefits, $f(x)$ on the other hand, need to be equally precise, nothing vague, such as life expectancy differential, some index of health, and similar quantities. If one cannot express the response quantitatively, then such an analysis cannot apply.

\subsection{ Antifragility}

\begin{figure} 
\includegraphics[width=\linewidth]{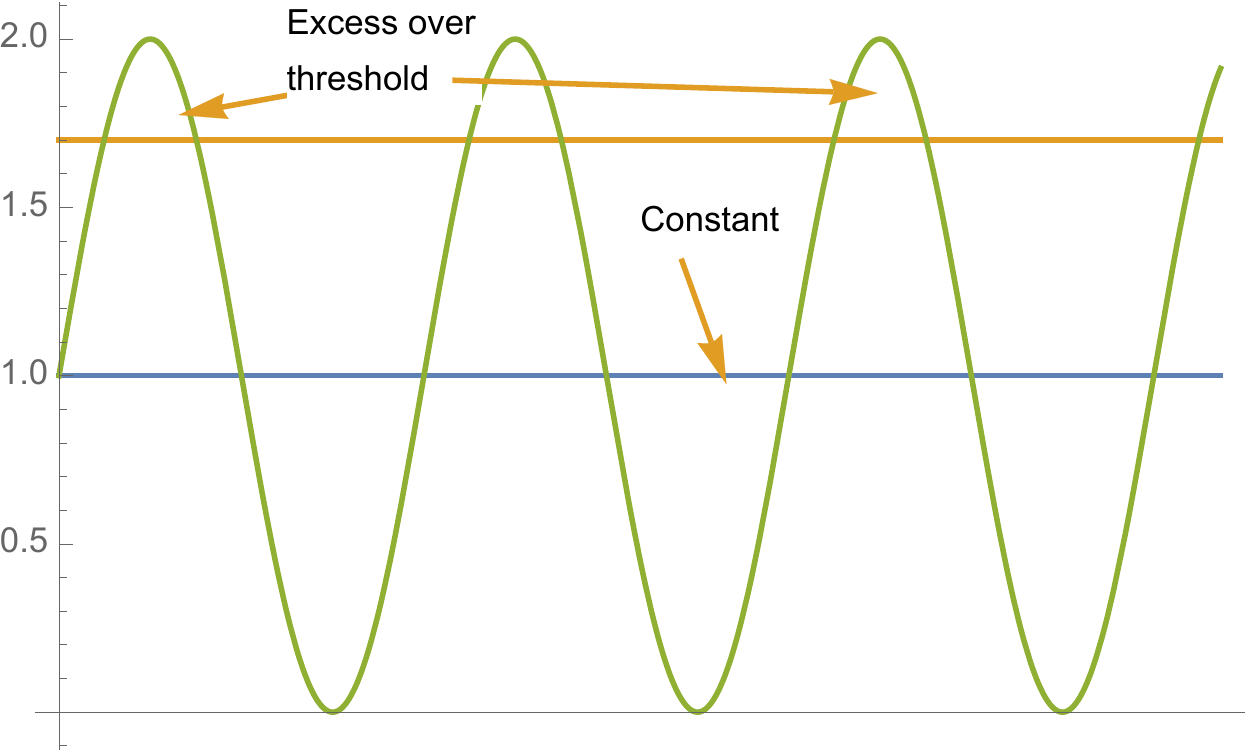}	
\caption{The figure shows why fractional intervention can be more effective in exceeding a threshold than constant dosage. This effect is similar to \textbf{stochastic resonance} known in physics by which noise cause signals to rise above the threshold of detection. For instance, genetically modified BT crops produce a constant level of pesticide, which appears to be much less effective than occasional manual interventions to add doses to conventional plants. The same may apply to antibiotics, chemotherapy and radiation therapy.}\label{dr3}
\end{figure}

\begin{figure} 
\includegraphics[width=\linewidth]{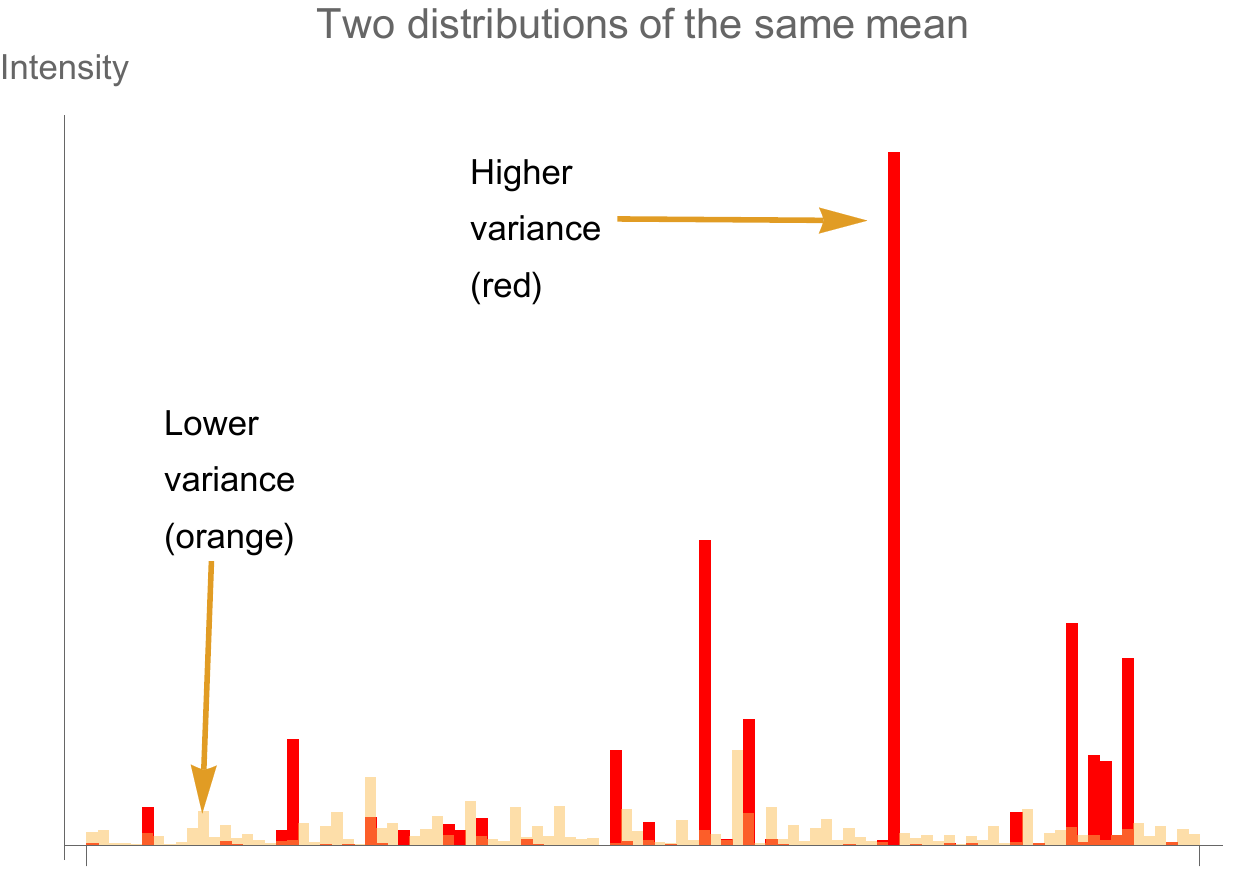}
\caption{An illustration of how a higher variance (hence scale), given the same mean, allow more spikes --hence an antifragile effect. We have a Monte Carlo simulations of two gamma distribution of same mean, different variances, $X_1\sim G(1,1)$ and $X_2\sim G(\frac{1}{10},10)$, showing higher spikes and maxima for $X_2$. The effect depends on norm $||.||_\infty $ , more sensitive to tail events, even more than just the scale which is  related to the norm $||.||_2 $.  }\label{dr4}
\includegraphics[width=\linewidth]{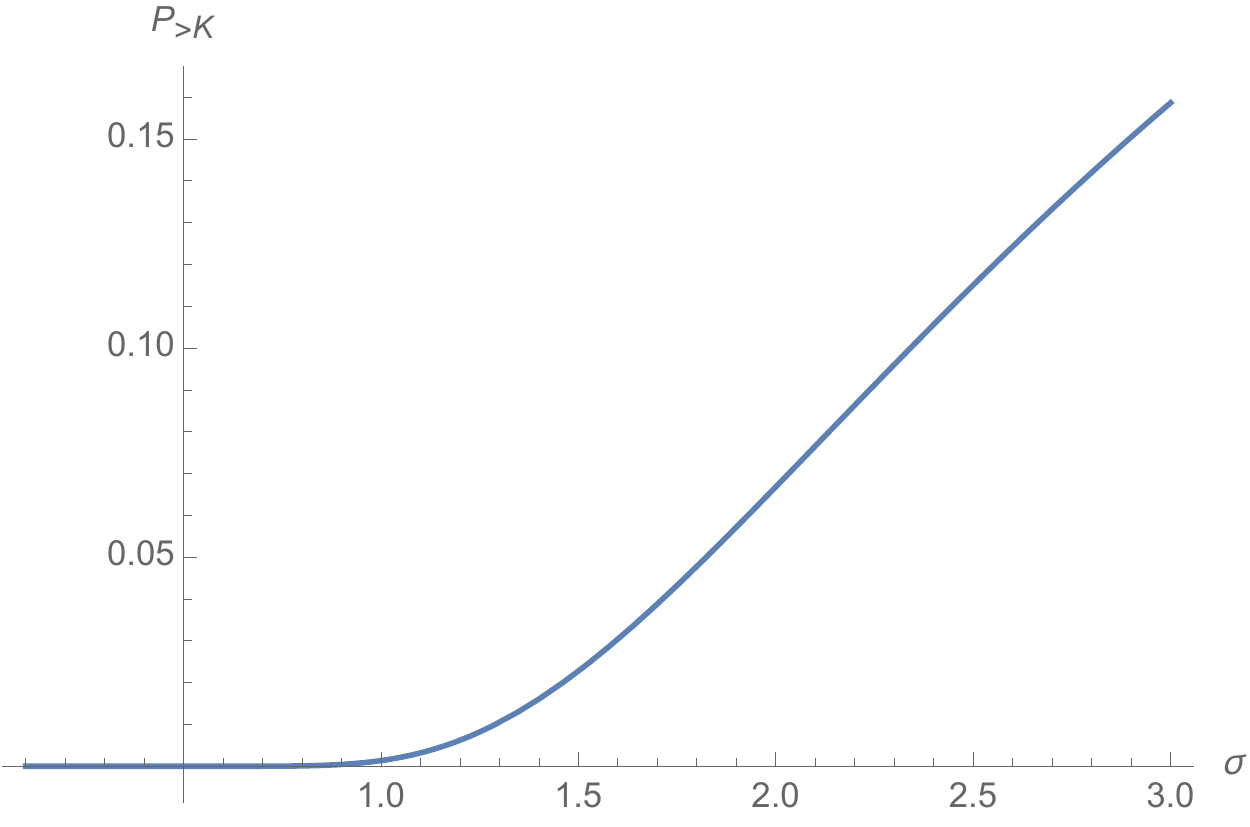}
\caption{Representation of Antifragility of Fig. \ref{dr4} in distribution space: we show the probability of exceeding a certain threshold for a variable, as a function of $\sigma$ the scale of the distribution, while keeping the mean constant.  }\label{dr5}
\end{figure}

\begin{figure} [h!]
\includegraphics[width=\linewidth]{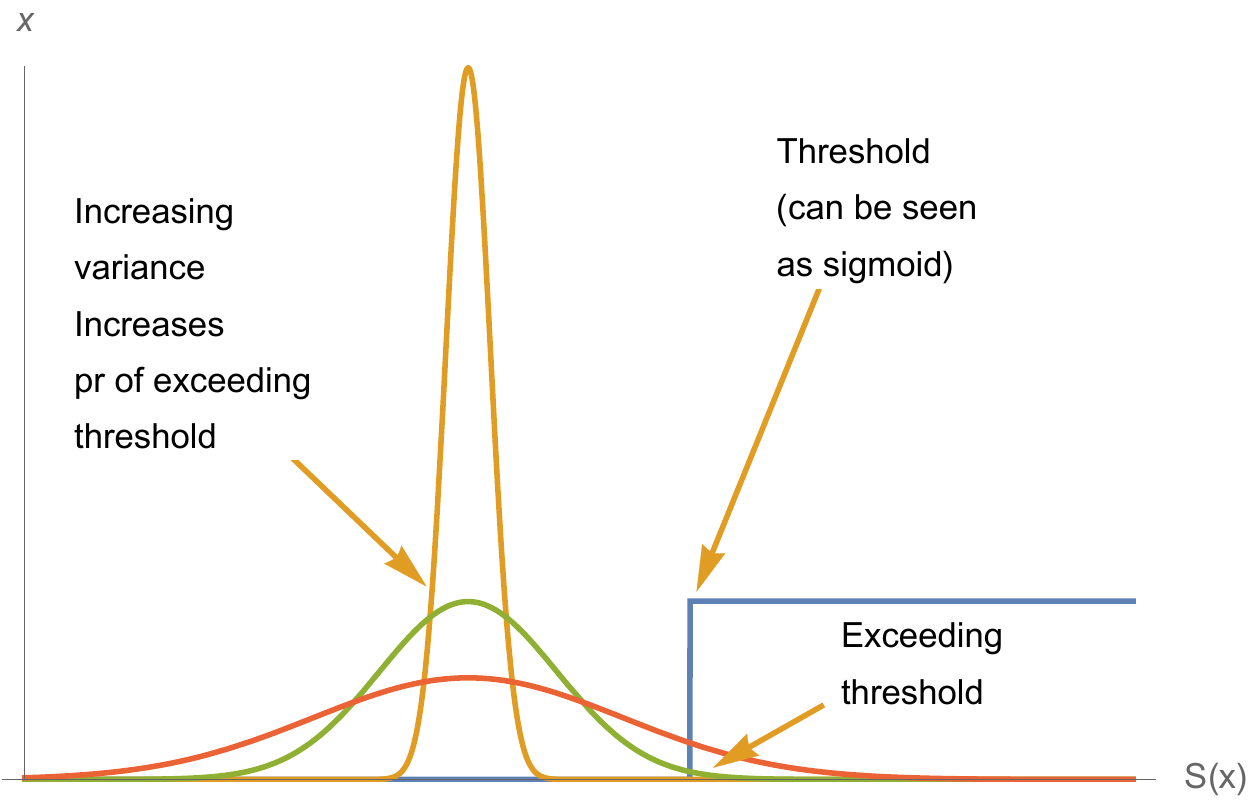}
\caption{How an increase in variance affects the threshold. If the threshold is above the mean, then we are in the presence of convexity and variance increases expected payoff more than changes in the mean, in proportion of the remoteness of the threshold. Note that the tails can be flipped (substituting the left for the right side) for the harm function if it is defined as negative.}\label{dr6}
\end{figure}

We define as locally antifragile\footnote{The term antifragile was coined in Taleb (2012) \cite{taleb2012antifragile} inspired from mathematical finance and derivatives trading, by which some payoff functions respond positively to increase in volatility and other measures of variation, a term in the vernacular called "long gamma".} a situation in which, over a specific interval $[a,b]$,  either the expectation increases with the scale of the distribution as in Eq. \ref{antifr}, or the dose response is convex over the same interval. The term in Taleb (2012) \cite{taleb2012antifragile} was meant to describe such a situation with precision: any situation that benefits from an increase in randomness or variability (since $\sigma$, the scale of the distribution, represents both); it is meant to be more precise than the vague "resilient" and bundle behaviors that "like" variability or spikes. 
Fig. \ref{dr3}, \ref{dr4}, \ref{dr5} and \ref{dr6} describe the threshold effect on the nonlinear response, and illustrates how they qualify as antifragile.

\subsection{ The first order sigmoid curve}
Define the sigmoid or sigmoidal function as having membership in a class of function $\mathfrak{S}$, $S: \mathbb{R}\rightarrow [L,H]$, with additional membership in the $ \mathbb{C}^2$ class (twice differentiable), monotonic nonincreasing or nondecreasing, that is let $S'(x)$ be the first derivative with respect to $x$:  $S'(x) \geq 0$  for all $x$ or $S'(x) \leq 0$. We have:
$$ S(x) = \left\{ \begin{array}{ll}
         H & \mbox{as $x \rightarrow +\infty$};\\
        L & \mbox{if $x \rightarrow -\infty$}.\end{array} \right., $$
        which can of course be normalized with $H=1$ and $L=0$ if $S$ is increasing, or vice versa, or alternatively $H=0$ and $L=-1$ if $S$ is increasing. We can define the simple (or first order) sigmoid curve as having equal convexity in one portion and concavity in another: $\exists k >0 \text{ s.t. } \forall x_1<k \text{ and } x_2>k,$ $\text{sgn} \left(S''(x_1)\right) =-\text{sgn}(S''(x_2))$ if   $|S''(x_2)|\geq 0$.
        
\begin{figure} 

	\includegraphics[width=\linewidth]{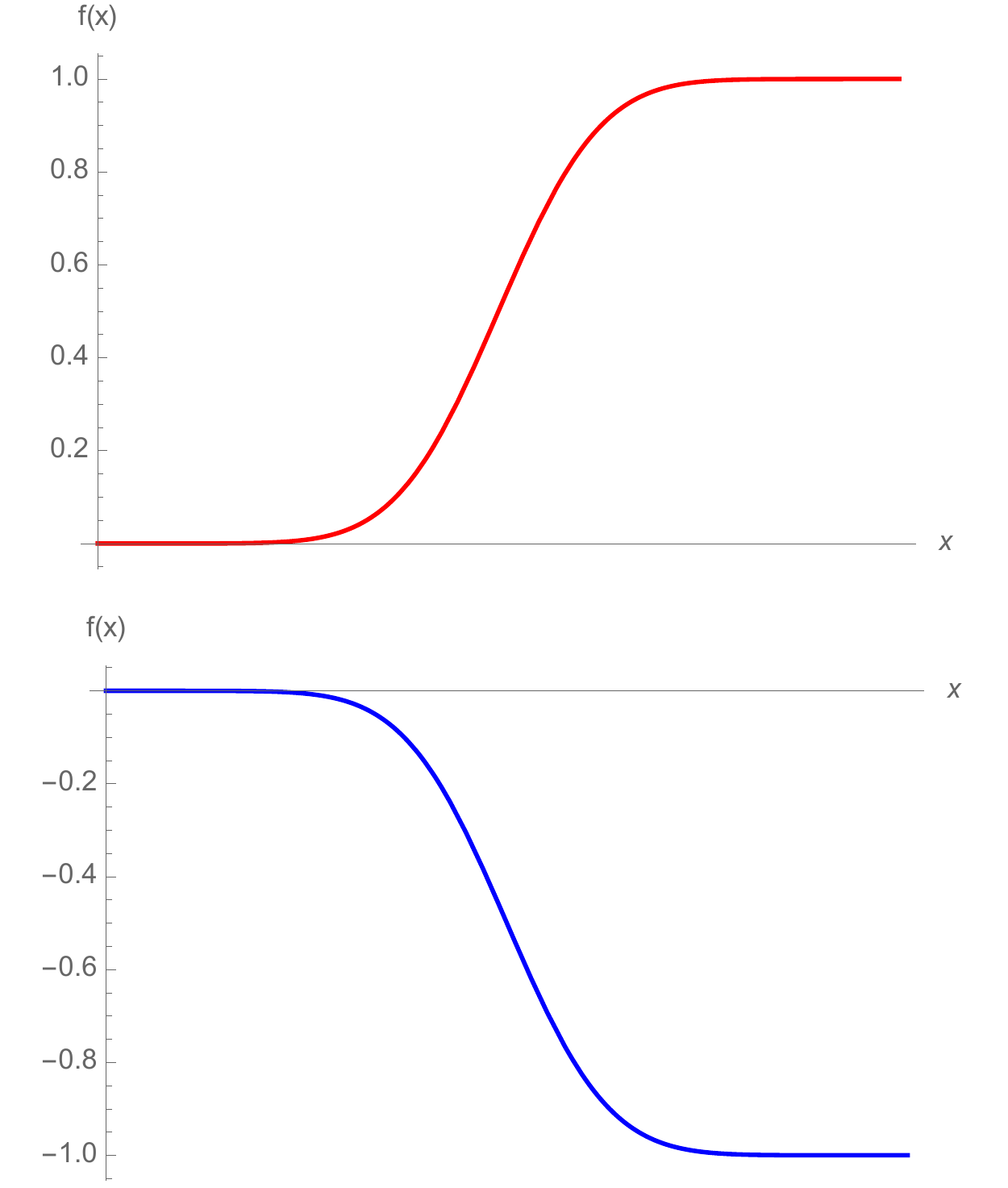}
	\caption{Simple (first order) nonincreasing or nondecreasing sigmoids} \label{simplesigmoids}
\end{figure}  

\begin{figure} 
	\includegraphics[width=\linewidth]{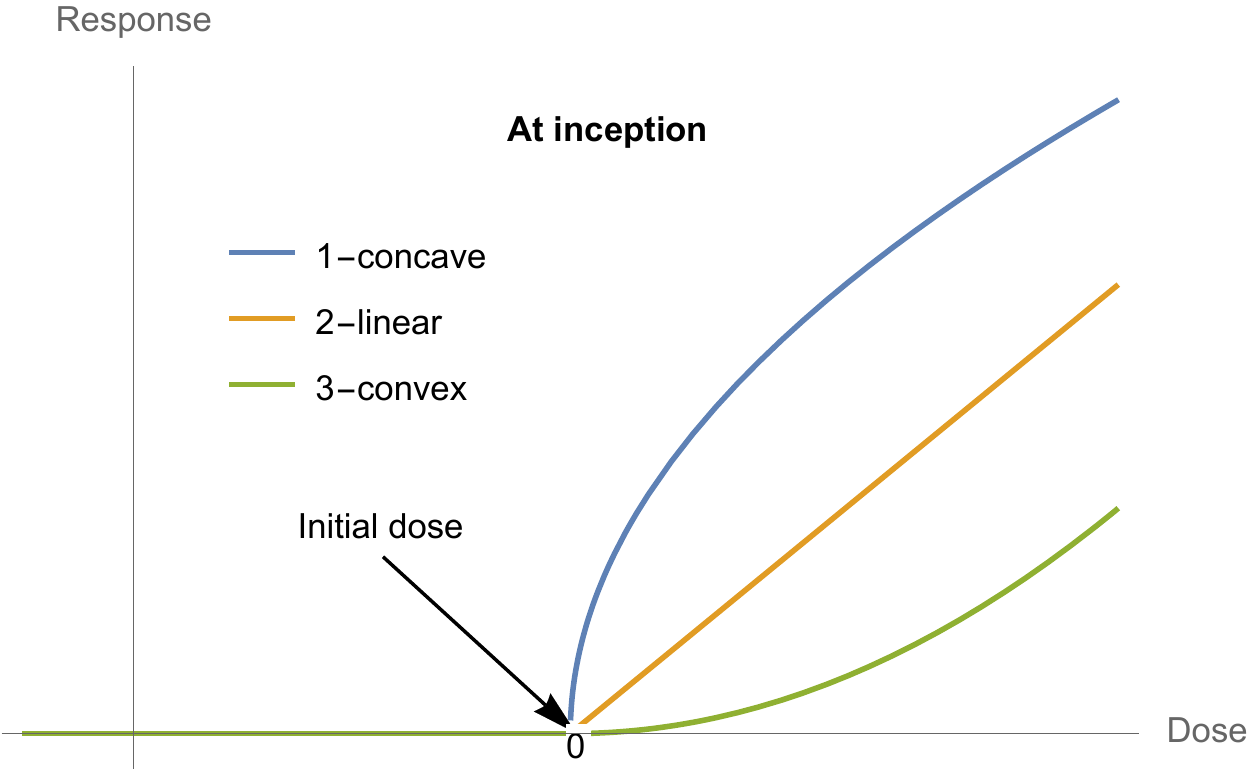}
	\caption{The three possibilities at inception} \label{threecurves}
	\end{figure}
\bigskip

Now all functions starting at 0 will have three possible properties at inception, as in Fig. \ref{threecurves}:
\begin{itemize}
	\item concave
	\item linear
	\item convex
\end{itemize}
The point of our discussion is the latter becomes sigmoid and is of interest to us. Although few medical examples appear, under scrutiny, to belong to the first two cases, one cannot exclude them from analysis. We note that given that the inception of these curves is 0, no linear combination can be initially convex unless the curve is convex, which would not be the case if the start of the reaction is at level different from 0.
 
	\begin{figure} 
	\includegraphics[width=\linewidth]{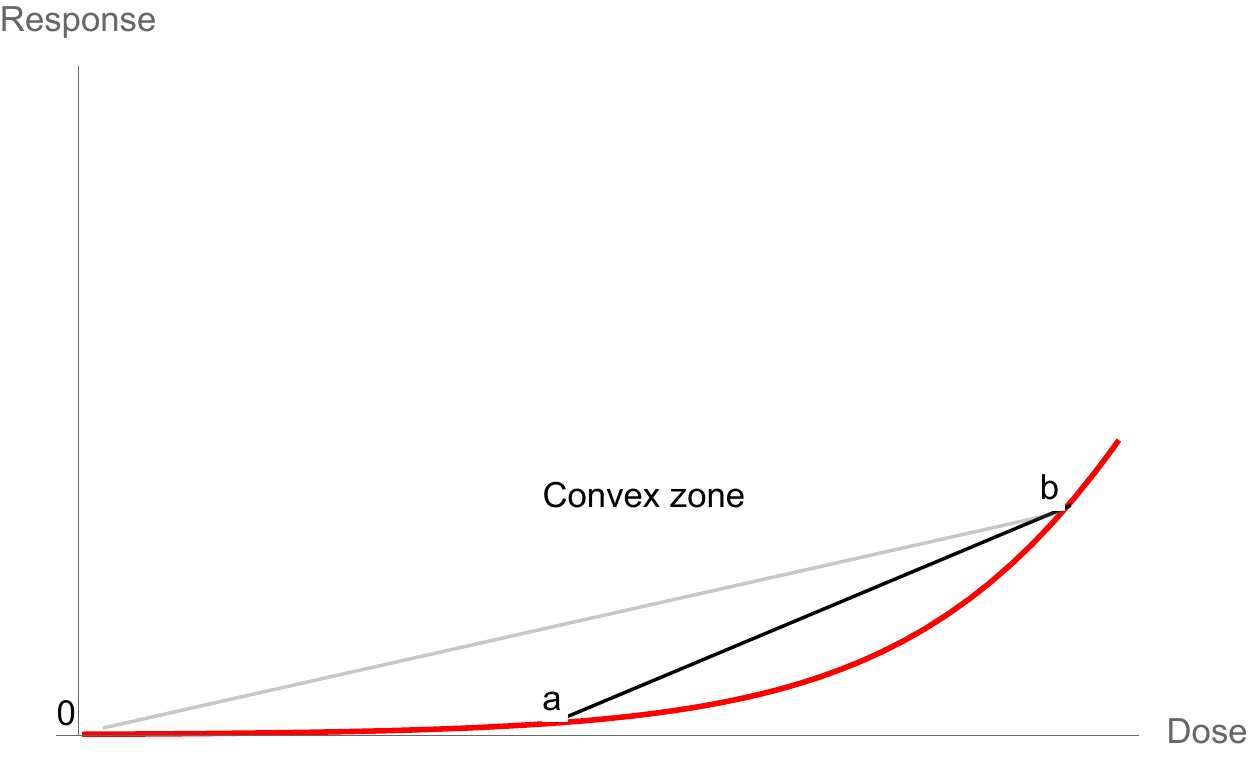}
	\caption{Every (relatively) smooth dose-response with a floor has to be convex, hence prefers variations and concentration} \label{dr1}
	\includegraphics[width=\linewidth]{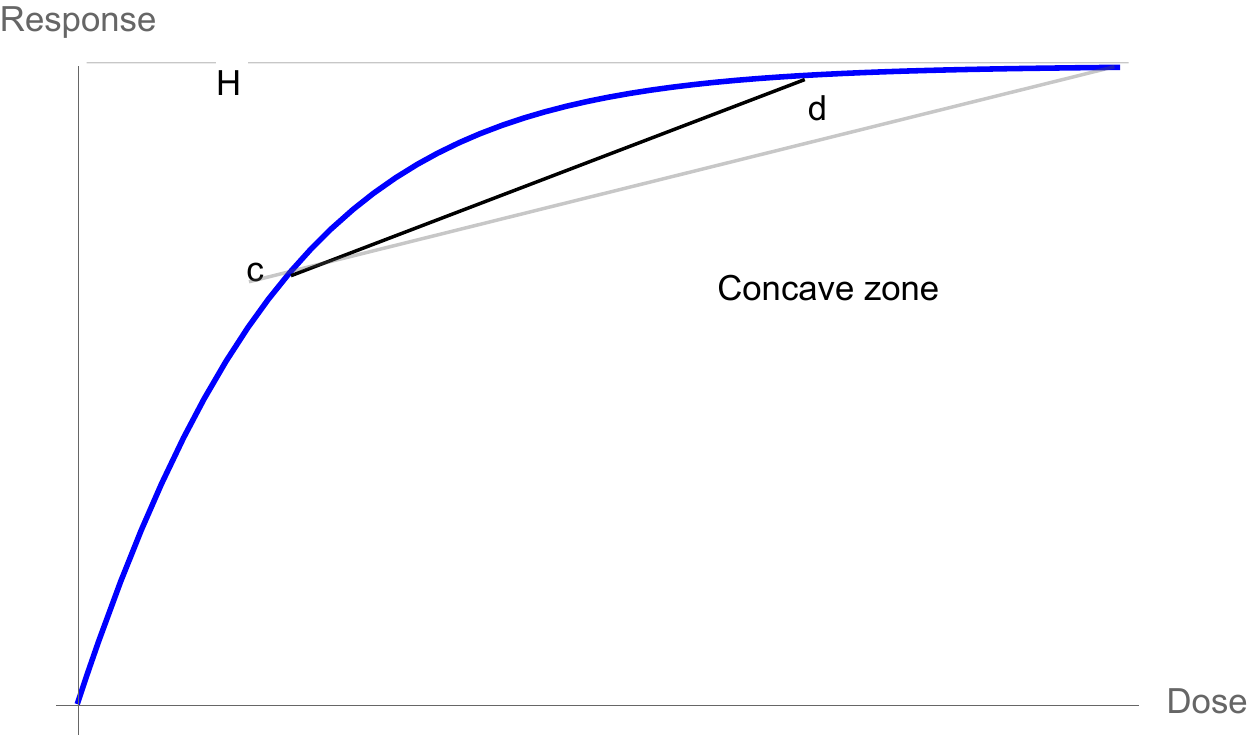}
	\caption{Every (relatively) smooth dose-response with a ceiling has to be concave, hence prefers stability}  \label{dr2}
\end{figure}

\begin{figure} [h!]
\includegraphics[width=\linewidth]{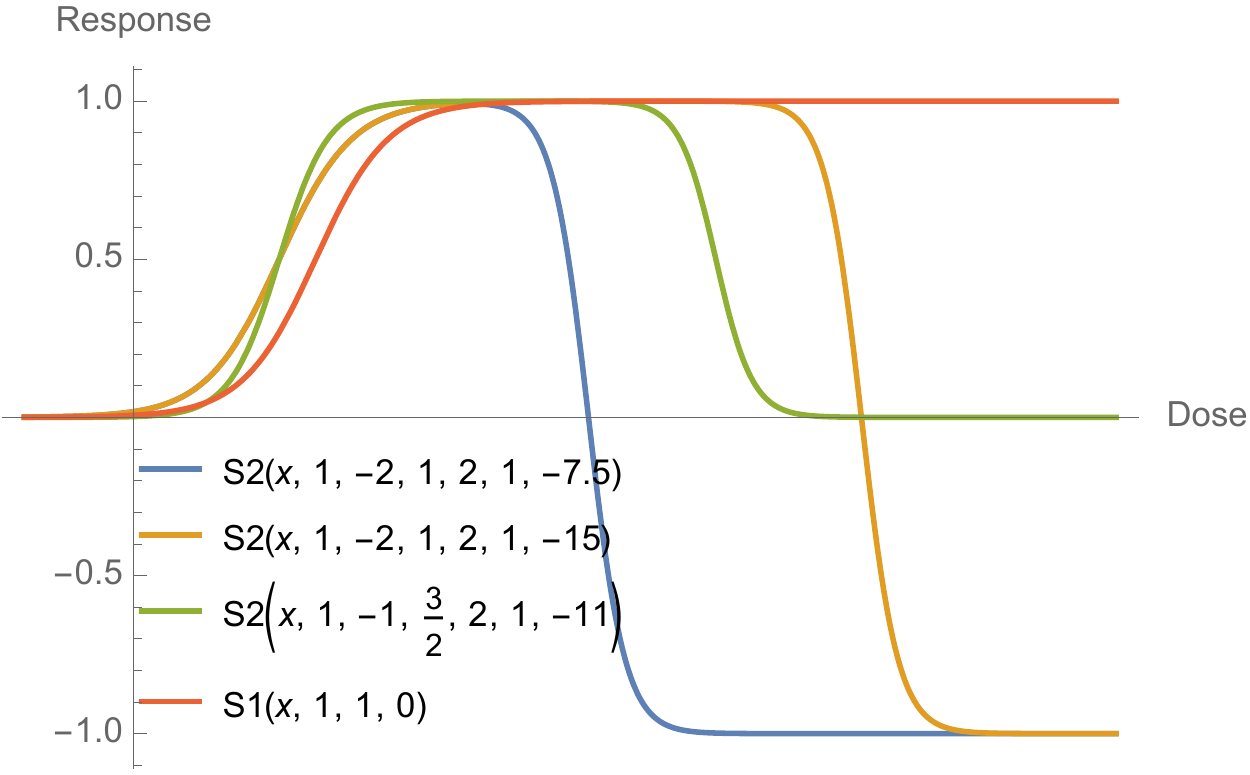}
\caption{The Generalized Response Curve, $S^2\left(x;a_1,a_2,b_1,b_2,c_1,c_2\right)\text{, }S^1\left(x;a_1,b_1,c_1\right)$ The convex part with positive first derivative has been designated as  "antifragile\index{Antifragility}"}\label{GDRCgraph}
\end{figure}[h!]

There are many sub-classes of functions producing a sigmoidal effect. Examples: 
\begin{itemize}
	\item Pure sigmoids with smoothness characteristics expressed in trigonometric or exponential form, $f: \mathbb{R} \rightarrow [0,1]$: 
	$$f(x)=\frac{1}{2} \tanh \left(\frac{\kappa  x}{\pi }\right)+\frac{1}{2}$$ $$f(x)=\frac{1}{1-e^{-a x}}$$
	\item Gompertz functions (a vague classification that includes above curves but can also mean special functions )
	\item Special functions with support in $\mathbb{R}$ such as the Error function 
	$f: \mathbb{R} \rightarrow [0,1]$
	$$f(x)=-\frac{1}{2} \text{erfc}\left(-\frac{x}{\sqrt{2}}\right)$$
	
	\item Special functions with support in $[0,1]$, such as  $f: [0,1] \rightarrow [0,1]$	$$f(x)=I_x(a,b),$$
where $I_{(.)}(.,.) $ is the Beta regularized function.
    \item Special functions with support in $[0,\infty)$
   $$f(x)=Q\left(a,0,\frac{x}{b}\right)$$
   where $Q\left(.,.,.\right) $ is the gamma regularized function.   
	\item Piecewise sigmoids, such as the CDF of the Student Distribution
	$$ f(x)=\begin{cases}
 \frac{1}{2} I_{\frac{\alpha }{x^2+\alpha }}\left(\frac{\alpha
   }{2},\frac{1}{2}\right) & x\leq 0 \\
 \frac{1}{2} \left(I_{\frac{x^2}{x^2+\alpha }}\left(\frac{1}{2},\frac{\alpha
   }{2}\right)+1\right) & x> 0
\end{cases}$$

\end{itemize}

\begin{figure} [h!]
	\includegraphics[width=\linewidth]{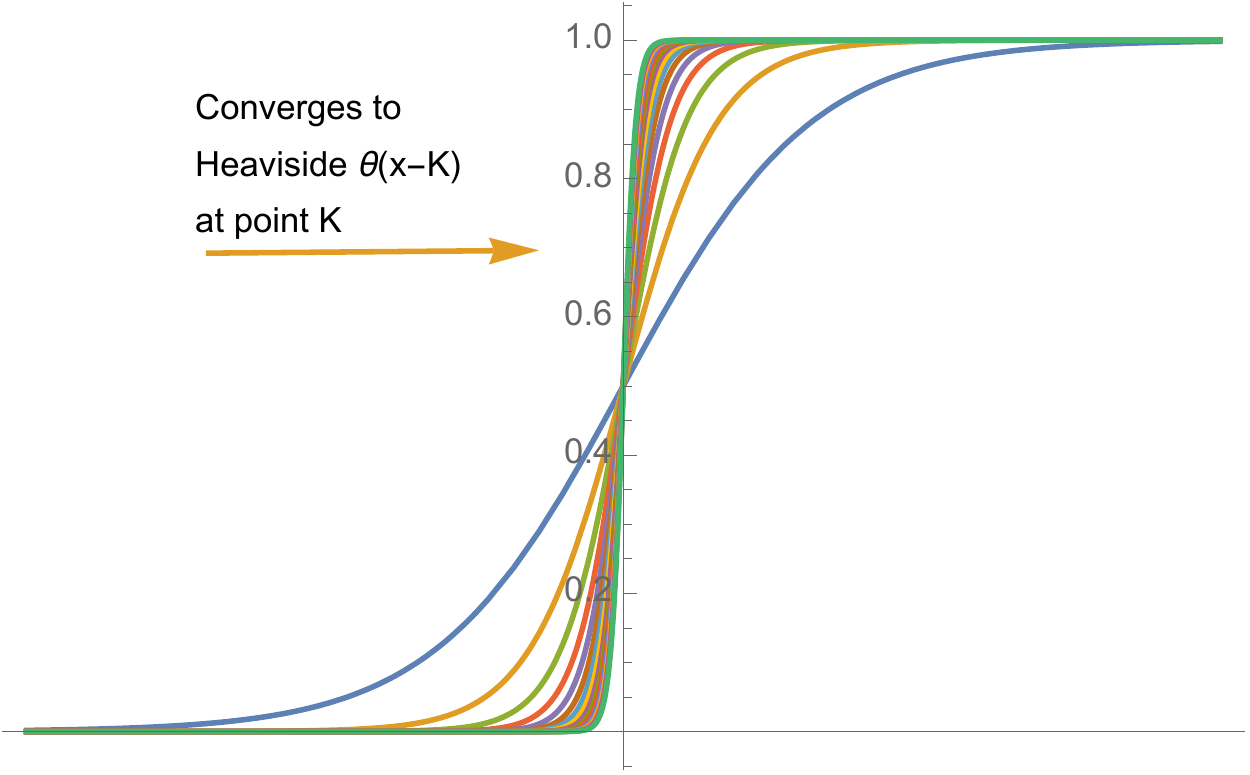}
	\caption{The smoothing of Heaviside as distribution or Schwartz function; we can treat step functions as sigmoid so long as $K$, the point of the step, is different from origin or endpoint.} \label{heaviside}
	\end{figure}

We note that the "smoothing" of the step function, or Heaviside theta $\theta(.)$ produces to a sigmoid (in a situation of a distribution  or convoluted with a test function with compact support), such as $\frac{1}{2} \tanh \left(\frac{\kappa  x}{\pi }\right)+\frac{1}{2}$, with $\kappa  \rightarrow \infty$, see Fig. \ref{heaviside}.

\subsection{ Some necessary relations leading to a sigmoid curve}

	Let $f_1(x): \mathbb{R}^+\rightarrow [0,H]$ , $H\geq 0$, of class $C^2$ be the first order dose-response function, satisfying $f_1(0)=0$, $f_1'(0)|=0$, $\lim_{x \to +\infty} f_1(x)=H$, monotonic nondecreasing, that is, $f_1'(x)\geq 0 \;\forall x \in \mathbb{R}^+$, with a continuous second derivative, and analytic in the vicinity of $0$. Then we conjecture that: 	
	
	\textbf{A}- There is exist a zone $[0,b]$ in which $f_1(x)$ is convex, that is $f_1''(x)\geq 0$, with the implication that $\forall a \leq b$ a policy of variation of dosage produces beneficial effects: $$\alpha f_1(a) +(1-\alpha) f_1(b)\geq f_1(\alpha a+ (1-\alpha) b), 0\leq \alpha\leq 1.$$
	(The acute outperforms the chronic).
	
	\textbf{B}- There is exist a zone $[c,H]$ in which $f_1(x)$ is concave, that is $f_1''(x)\leq 0$, with the implication that $\exists d \geq c$ a policy of stability of dosage produces beneficial effects: $$\alpha f_1(c) +(1-\alpha) f_1(d)\leq f_1(\alpha c+ (1-\alpha) d).$$
	(The chronic outperforms the acute).
%

\section{ The Generalized Dose Response Curve}\label{GDRC}

Let $S^N(x)$: $\mathbb{R}$ $\to $ [$k_L$, $k_R$], $S^N \in C^{\infty}$, be a continuous function possessing derivatives $\left(S^N\right)^{(n)}(x)$ of all orders, expressed as an $N$-summed and scaled standard sigmoid  functions:

\begin{equation}
S^N(x) \triangleq  \sum _{i=1}^N \frac{a_k}{1+e^{\left(-b_k x+c_k\right)}}\label{gensig}
\end{equation}

where $a_k, b_k, c_k$ are scaling constants $\in $ $\mathbb{R}$, satisfying: 

i) $S^N$(-$\infty $) =$k_L$ 

ii) $S^N$($+\infty $) =$k_R$

and (equivalently for the first and last of the following conditions)

iii) $\frac{\partial ^2 S^N}{\partial x^2}$$\geq $ 0 { }for $x$ $\in $ (-$\infty $, $k_1$) , $\frac{\partial ^2 S^N}{\partial x^2}$$<$ 0 for $x$ $\in $ ($k_2$, $k_{>2}$), and $\frac{\partial ^2 S^N}{\partial x^2}$$\geq $ 0 for $x$ $\in $ ($k_{>2}$, $\infty $), 
with $k_1>k_2\geq \ldots \geq k_N$.

By increasing $N$, we can approximate a continuous functions dense in a metric space, see Cybenko (1989) \cite{cybenko1989approximation}.

The shapes at different calibrations are shown in Fig. \ref{GDRCgraph}, in which we combined different values of N=2 $S^2\left(x;a_1,a_2,b_1,b_2,c_1,c_2\right)\text{, }$and the standard sigmoid  $S^1\left(x;a_1,b_1,c_1\right)$, with $a_1$=1, $b_1$=1 and $c_1$=0. As we can see, unlike the common sigmoid , the asymptotic response can be lower than the maximum, as our curves are not monotonically increasing. The sigmoid shows benefits increasing rapidly (the convex phase), then increasing at a slower and slower rate until saturation. Our more general case starts by increasing, but the reponse can be actually negative beyond the saturation phase, though in a convex manner. Harm slows down and becomes "flat" when something is totally broken.

\section{ Antifragility in the various literatures}

Before moving to the iatrogenics section, let us review the various literature that found benefits in increase in scale (i.e. local antifragility) though without gluing their results as part of a general function.

 In short the papers in this section show \textit{indirectly} the effects of an increase in $\sigma$ for diabetes, alzheimer, cancer rates, or whatever condition they studied. The scale of the distribution means increasing the variance, say instead of giving a feeding of $x$ over each time step $\Delta t$, giving $x -\delta$ then $x +\delta$ instead, as in Eqs. \ref{jensen} and \ref{antifr}.  Simply, intermittent fasting would be having $\Delta \approx x$. and the scale can be written in such a simplified example as the dispersion $\sigma \approx \delta$. 
 \bigskip
  
 \subsection{ Denial of second order effect}
 In short, antifragility is second order effect (the average is the first order effect).
 
 One blatant mistake in the literature lies in ignoring the second order effect when making statements from empirical data. An illustration is dietary recommendations based on composition without regard to frequency. For instance, the use of epidemiological data concerning the Cretan diet focused on composition and not how often people ate each food type. Yet frequency matters: the Greek Orthodox church has, depending on the severity of the local culture, almost two hundred vegan days per year, that is, an episodic protein deprivation; meats are eaten in lumps that compensate for the deprivation. As we will see with the literature below, there is a missing mathematical bridge between studies of \textit{variability}, say Mattson et al.(2006) and Fontana et al (2008) on one hand, and the focus on food \textit{composition} --the Longo and Fontana studies, furthermore, narrows the effect of the frequency to a given food type, namely proteins\footnote{Lee and Longo (2011) \cite{lee2011fasting} "In the prokaryote E. coli, lack of glucose or nitrogen (comparable to protein restriction in mammals) increase resistance to high levels of $H_2 O_2$ (15 mm) (Jenkins et al., 1988) \cite{Jenkins1988starvation}"}.
 
 Further, the computation of the "recommended daily" units may vary markedly if one assumes second order effects: the needed average is mathematically sensitive to frequency, as we saw earlier.

 \subsection{ Scouring the literature for antifragility}
A sample of papers document such reaction to $\sigma$ is as follows.

\begin{figure} [h!]
	\includegraphics[width=\linewidth]{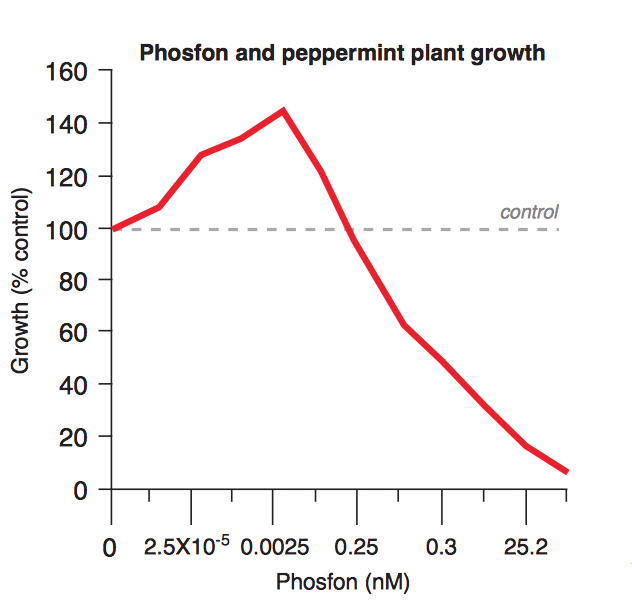}
\caption{ Hormesis in Kaiser (2003) we can detext a convex-concave sigmoidal shape that fits our generalized sigmoid in Eq.\ref{gensig}.}\label{kaiser}
\end{figure}

Mithridatization and hormesis:  Kaiser (2003) \cite{kaiser2003sipping} (see Fig. \ref{kaiser}), Rattan (2008) \cite{rattan2008hormesis}, Calabrese and Baldwin (2002, 2003a, 2003b) \cite{calabrese2002defining},\cite{calabrese2003hormesis},\cite{calabrese2003hormetic}, Aruguman et al (2006) \cite{arumugam2006hormesis}. Note that the literature focuses on mechanisms and misses the explicit convexity argument. Is also absent the idea of divergence from, or convergence to the norm --hormesis might just be reinstatement of normalcy as we will discuss further down. 

\paragraph{Caloric restriction and hormesis:} Martin, Mattson et al. (2006) \cite{martin2006caloric}.

\paragraph{Treatment of various diseases:} Longo and Mattson(2014) \cite{longo2014fasting}. 
\paragraph{Cancer treatment and fasting:} Longo et al. (2010) \cite{longo2010calorie}, Safdie et al. (2009) \cite{safdie2009fasting}, Raffaghelo et al. (2010), \cite{raffaghello2010fasting}, Lee et al (2012) \cite{lee2012fasting}.

\paragraph{Aging and intermittence:} Fontana et al. \cite{fontana2014medical}.
\paragraph{For brain effects:}
Anson, Guo, et al. (2003) \cite{anson2003intermittent}, Halagappa, Guo, et al. (2007) \cite{halagappa2007intermittent}, Stranahan and Mattson (2012) \cite{stranahan2012recruiting}. The long-held belief that the brain needed glucose, not ketones, and that the brain does not go through autophagy, has been progressively replaced.
\paragraph{On yeast and longevity under restriction;} Fabrizio et al. (2001)\cite{fabrizio2001regulation}; SIRT1, Longo et al. (2006) \cite{longo2006sirtuins}, Michan et al. (2010) \cite{michan2010sirt1}.

 \paragraph{For diabetes, remission or reversal:} Taylor (2008) \cite{taylor2008pathogenesis}, Lim et al. (2011) \cite{lim2011reversal}, Boucher et al. (2004) \cite{boucher2004biochemical}; diabetes management by diet alone, early insights in Wilson et al. (1980) \cite{wilson1980dietary}.  Couzin (2008) \cite{couzin2008deaths} gives insight that blood sugar stabilization does not have the effect anticipated (which in our language implies that $\sigma$ matters). 
The ACCORD study (Action to Control Cardiovascular Risk in Diabetes) found no gain from lowering blood glucose, or other metrics --indeed, it may be more opaque than a simple glucose problem remedied by pharmacological means. Synthesis, Skyler et al. (2009) \cite{skyler2009intensive}, old methods, Westman and Vernon (2008) \cite{westman2008has}.
Bariatric (or other) surgery as an alternative approach from intermittent fasting: Pories (1995) \cite{pories1995would}, Guidone et al. (2006) \cite{guidone2006mechanisms}, Rubino et al. 2006 \cite{rubino2006mechanism}. 
  
\paragraph{Ramadan and effect of fasting:} Trabelsi et al. (2012) \cite{trabelsi2012effect}, Akanji et al. (2012). Note that the Ramadan time window is short (12 to 17 hours) and possibly fraught with overeating so conclusions need to take into account energy balance and that the considered effect is at the low-frequency part of the timescale.


\paragraph{Caloric restriction:} Harrison (1984), Wiendruch (1996), Pischon (2008). An understanding of such natural antifragility  can allow us to dispense with the far more speculative approach of pharmalogical interventions such as suggested in Stip (2010) --owing to more iatrogenics discussed in the next section\ref{iatrosection}.
\paragraph{Autophagy for cancer:} Kondo et al. (2005) \cite{Kondo2005role}.

\paragraph{Autophagy (general):} Danchin et al. (2011) \cite{danchin2011antifragility}, He et al. (2012) \cite{he2012exercise}.

\paragraph{Fractional dosage:} Wu et al. (2016) \cite{wu2016fractional}.
Jensen's inequality in workout: Many such as Schnohr and Marott (2011) \cite{schnohr2011intensity} compare the results of intermittent extremes with "moderate" physical activity; they
got close to dealing with the fact that extreme sprinting and nothing  outperforms steady exercise, but missed the convexity bias part.

\paragraph{Cluster of ailments:} Yaffe and Blackwell (2004) \cite{yaffe2004diabetes}, Alzheimer and hyperinsulenemia as correlated, Razay and Wilcock (1994) \cite{razay1994hyperinsulinaemia};  Luchsinger, Tang, et al. (2002) \cite{luchsinger2002caloric}, Luchsinger Tang et al. (2004) \cite{luchsinger2004hyperinsulinemia} Janson, Laedtke, et al. (2004) \cite{janson2004increased}. The clusters are of special interest as they indicate how the absence or presence of convex effect can manifest itself in multiple diseases.

\paragraph{Benefits  of \textit{some type of} stress (and convexity of the effect): }For the different results from the two types of stressors, short and chronic, Dhabar (2009) "A hassle a day may keep the pathogens away: the fight-or-flight stress response and the augmentation of immune function" \cite{dhabhar2009hassle}.  for the benefits of stress on boosting immunity and cancer resistance (squamous cell carcinoma), Dhabhar et al. (2010) \cite{dhabhar2010short}, Dhabhar et al. (2012) \cite{dhabhar2012high} , Ansbacher et al. (2013)\cite{aschbacher2013good}.

\paragraph{Iatrogenics of hygiene and systematic elimination of germs:} Rook (2011) \cite{rook2011hygiene}, Rook (2012) \cite{rook2012hygiene} (auto-immune diseases from absence of stressor), M\'egraud and Lamouliatte (1992) \cite{megraud1992helicobacter} for Helyobacter Pilori and incidence of cancer.

\subsection{ Extracting an ancestral frequency}
We noted that papers such as Kaiser (2003) \cite{kaiser2003sipping} and Calabrese and Baldwin (2003) , \cite{calabrese2003hormesis} miss the point that hormesis may correspond to a "fitness dose", beyond and below which one departs from such ideal dispersion of the dose $x$ per time period. 

 We can also apply the visible dose-response curve to inferring the ideal parametrization of the probability distribution for our feeding (ancestral or otherwise) and vice-versa. For instance, measuring the effects of episodic fasting on cancer, diabetes, and other ailments can lead to assessing some kind of "fitness" to an environment with a certain structure of randomness, either with the $\sigma$ above or some richer measure of probability distribution. Simply, if diabetes can be controlled or reversed with occasional deprivation (a certain variance), say 24 hour fasts per week, 3 days per quarter, and a full week every four years, then necessarily our system can be made to fit stochastic energy supply, with a certain frequency of deficits --and, crucially, we can extract the functional expression from such frequencies.
 \bigskip
  
	Note that an understanding of the precise mechanism by which intermittence works (whether dietary or in energy expenditure), which can be autophagy or some other mechanism such as insulin control, are helpful but not needed given the robustness of the mathematical link between the functional and the probabilistic.

\section{ Nonlinearities and Iatrogenics}\label{iatrosection}
\begin{figure} 
	\includegraphics[width=\linewidth]{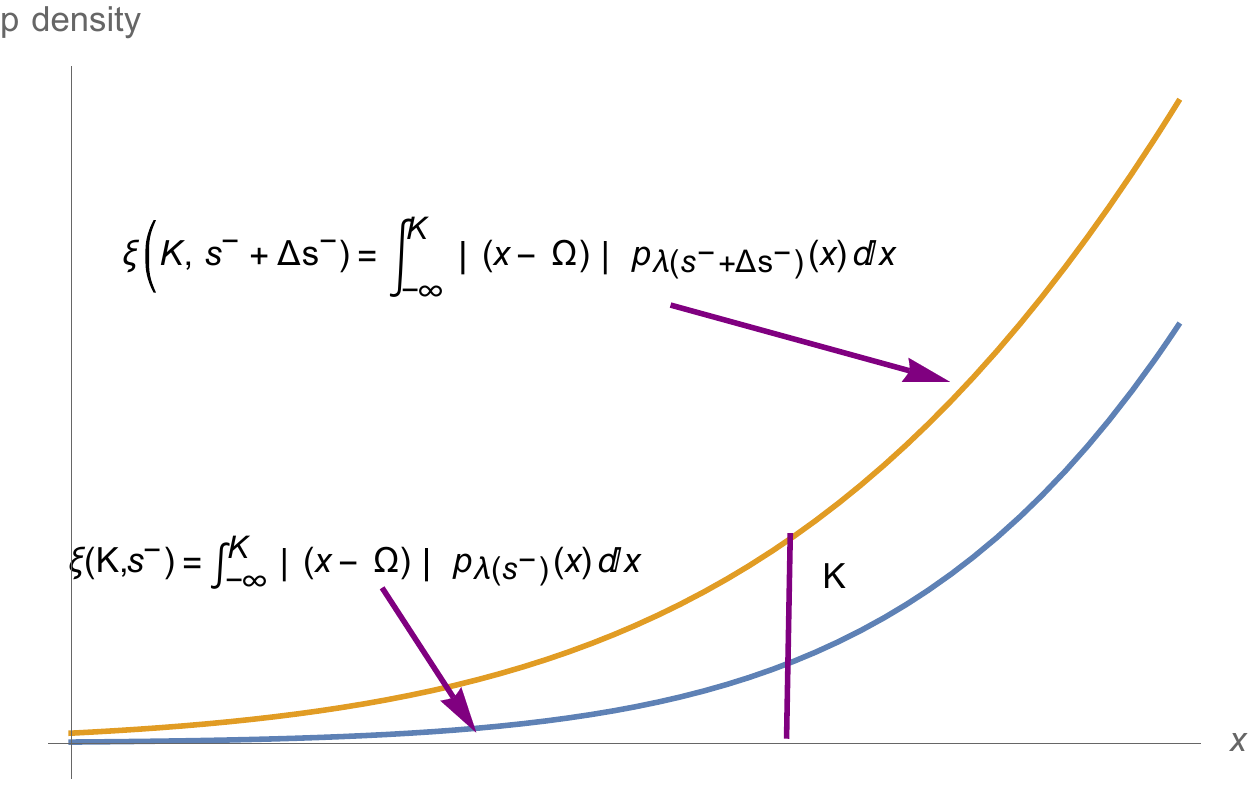}
\caption{ A definition of fragility as left tail payoff sensitivity; the figure shows the effect of the perturbation of the lower semi-deviation $s^-$ on the tail integral ${\xi}$ of
$(x -\Omega)$ below $K$, $\Omega$ being a centering constant. Our detection of fragility does not require the
specification of $p$ the probability distribution.}\label{tailvegagraph}
\end{figure}

Next we connect nonlinearity to iatrogenics, broadly defined as all manner of net deficit of benefits minus harm from a given intervention. 

In short, Taleb and Douady (2013) \cite{taleb2013mathematical} describes fragility as a "tail" property, that is, below a set level $K$, how either 1) greater uncertainty or 2) more variability translate into a degradation of the effect of the probability distribution on the expected payoff.
\bigskip
 
The probability distribution of concern has for density $p$, a scale $s^-$ for the distribution below $\Omega$ a centering constant (we can call $s^-$
a negative semideviation). To cover a broader set of distributions, we use $p_{\lambda(s)}$ where $\lambda$ is a function of $s$.

We set $\xi(.,.)$ a function of the expected value below $K$. Intuitively it is meant to express the harm, and, mostly its variations --one may not have a precise idea of the harm but the variations can be extracted in a more robust way.

\begin{equation}
\xi(s^-)=\int _{-\infty }^K|x-\Omega | \,p_{\lambda (s^-)(x)} \; \mathrm{d}x	
\end{equation}

\begin{equation}
\xi(s^-+\Delta s^-)=\int _{-\infty }^K|x-\Omega | \,p_{\lambda (s^-+\Delta s^-)} (x) \; \mathrm{d}x	
\end{equation}

\bigskip
 
Fragility is defined as the variations of $\xi(.)$ from an increase in the left scale $s^- $ as shown in Fig \ref{tailvegagraph}. The difference  $\xi(\Delta s^-)$ represents a sensitivity to an expansion in uncertainty in the left tail.

\bigskip

The theorems in Taleb and Douady (2013) \cite{taleb2013mathematical} show that: 
\begin{itemize}
	
\item Convexity in a dose-response function increases $\xi$.	

\item Detecting such nonlinearity allows us to predict fragility and formulate a probabilistic decision without knowing $p(.)$.

\item The mere existence of concavity in the tails implies an unseen risk.

\end{itemize}

\begin{figure} 
\includegraphics[width=\linewidth]{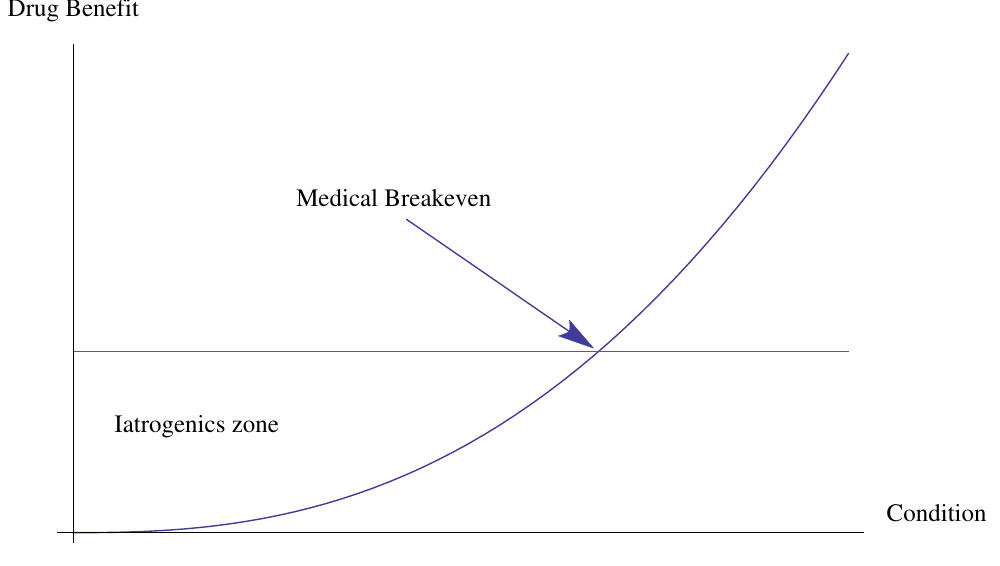}	
\caption{Drug benefits when convex to Numbers Needed to Treat, with gross iatrogenics invariant to condition (the constant line). We are looking at the convex portion of a possibly sigmoidal benefit function. }\label{nnt}
\end{figure}

\begin{figure} 
\includegraphics[width=\linewidth]{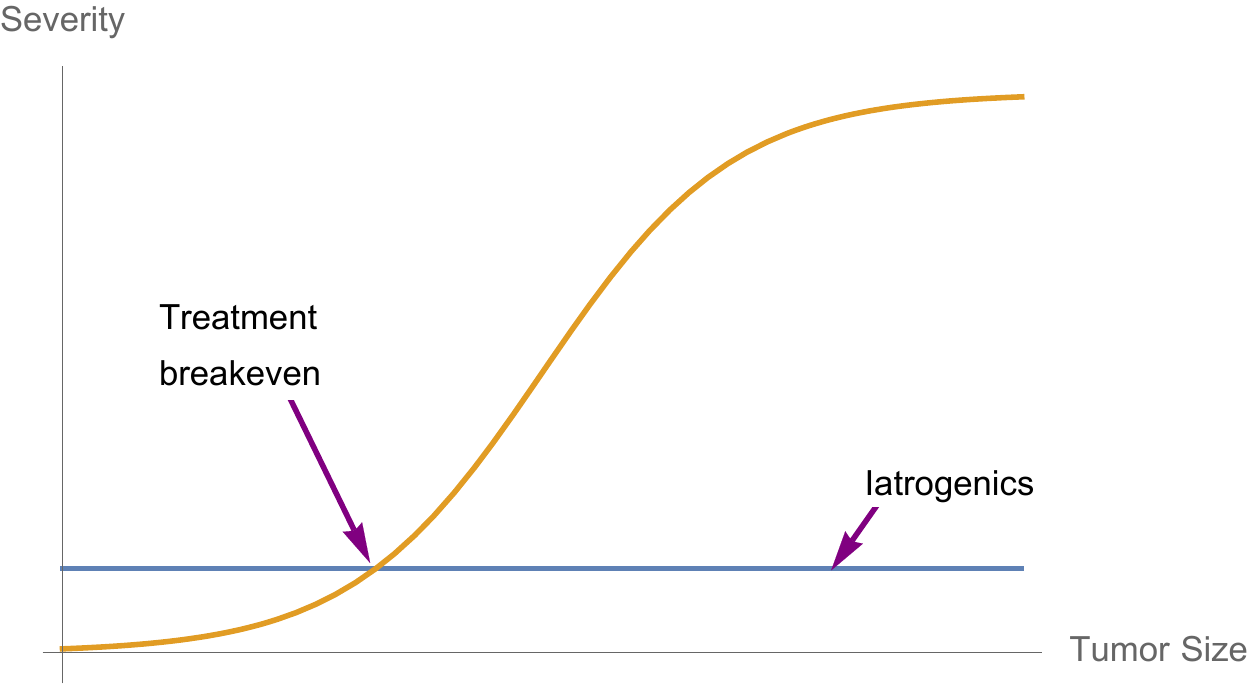}
\caption{Tumor breakeven we consider a wider range of Fig. \ref{nnt} and apply it to the relation between tumor size and treatment breakeven.}\label{nnt2}
\end{figure}

\begin{figure} 
\includegraphics[width=\linewidth]{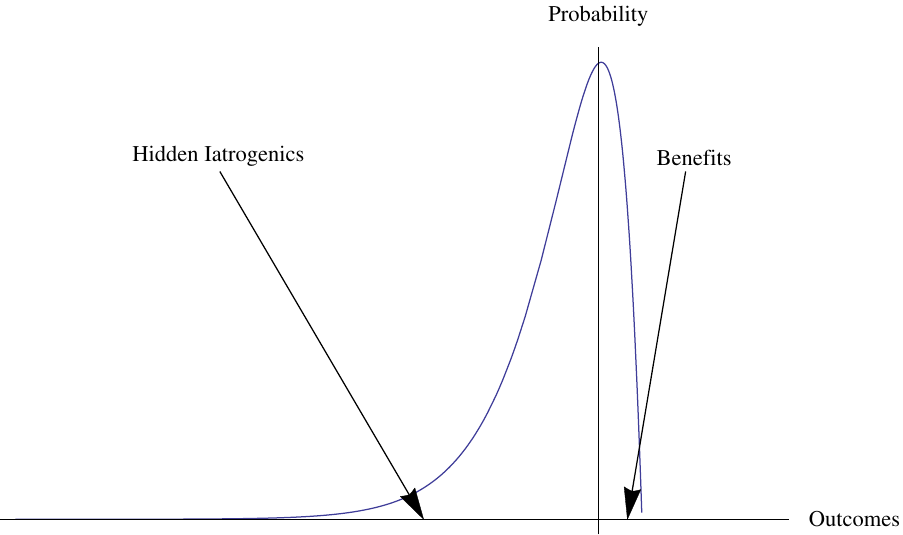}
\caption{Unseen risks and mild gains: translation of Fig. \ref{nnt} to the skewness of a decision involving iatrogenics when the condition is mild. This also illustrates the Taleb and Douady\cite{taleb2013mathematical} translation theorems from concavity for $S(x)$ into a probabilistic attributes.}\label{skewness}	
\end{figure}

\begin{figure}[h!]
\includegraphics[width=\linewidth]{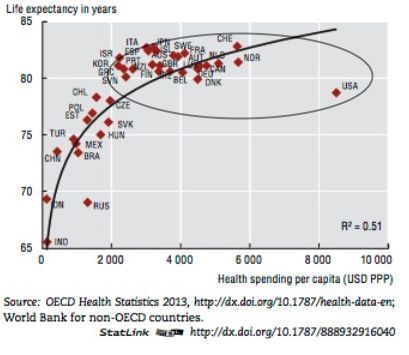}
\caption{Concavity of Gains to Health Spending. A more appropriate regression line than the one used by OECD should flatten off to the right, even invert to  fit the USA.  Credit: Edward Tufte}\label{healthstats}
\end{figure}

\subsection{ Effect reversal and the sigmoid}

Now let us discuss Figs. \ref{nnt} and \ref{nnt2}. The nonlinearities of dose response and hormetic or neutral effect at low doses is illustrated in the case of radiation: In Neumaier et al. (2012) \cite{neumaier2012evidence} titled "Evidence for formation of DNA repair centers and dose-response nonlinearity in human cells", the authors write: "The standard model currently in use applies a linear scale, extrapolating cancer risk from high doses to low doses of ionizing radiation. However, our discovery of DSB clustering over such large distances casts considerable doubts on the general assumption that risk to ionizing radiation is proportional to dose, and instead provides a mechanism that could more accurately address risk dose dependency of ionizing radiation." Radiation hormesis is the idea that low-level radiation causes hormetic overreaction with protective effects. Also see Tubiana et al. (2005) \cite{tubiana2006recent}.

Bharadwaj and Stafford (2010) present similar general-sigmoidal effects in hormonal disruptions by chemicals \cite{bharadwaj2010hormones}.

\subsection{ Nonlinearity of NNT, overtreatment, and decision-making}

Below are applications of convexity analysis in decision-making in dosage, shown in Fig. \ref{nnt}, \ref{nnt2} and Fig. \ref{skewness}. 
\bigskip
 
In short, it is fallacious to translate a policy derived from acute conditions and apply it to milder ones. Mild conditions are different in treatment from an acute one. 
	
Likewise, high risk is qualitatively different from mild risk.
 
\subsubsection{Mammogram controversy} There is an active literature on "overdiagnosis", see Kalager et al(2012) \cite{kalager2012overdiagnosis}, Morell et al.(2012) \cite{morrell2010estimates}. The point is that treating a tumor that doesn't kill reduces life expectancy; hence the need to balance iatrogenics and risk of cancer. An application of nonlinearity can shed some light to the approach, particularly that public opinion might find it "cruel" to deprive people of treatment even if it extends their life expectancy \cite{taleb2012antifragile}. 

\subsubsection{Hypertension illustrations}

Consider the following simplified case from blood pressure studies: assume that when hypertension is mild, say marginally higher than the zone accepted as normotensive, the chance of benefiting from a certain drug is close to 5\% (1 in 20). But when blood pressure is considered to be in the "high" or "severe" range, the chances of benefiting would now be 26\% and 72\%, respectively. See Pearce et al (1998) \cite{pearce1998cost} for similar results for near-nomotensive.

But consider that (unless one has a special reason against) the iatrogenics should be safely considered constant for all categories. In the very ill condition, the benefits are large relative to iatrogenics; in the borderline one, they are small. This means that we need to focus on high-symptom conditions compare to other situations in which the patient is not very ill. 

 A 2012 Cochrane meta-analysis indicated that there is no evidence that treating otherwise healthy mild hypertension patients with antihypertensive therapy will reduce CV events or mortality. Makridakis and DiNicolantonio (2014)   \cite{makridakis2014hypertension} found no statistical basis for current hypertension treatment.  Rosansky(2012)\cite{ rosansky2012hypertension} found a "silent killer" in iatrogenics, i.e. hidden risks, matching our illustration in distribution space in Fig. \ref{skewness}.

\subsubsection{Statin example}
We can apply the method to statins, which appears to have benefits in the very ill segment that do not translate into milder conditions.  With statin drugs routinely prescribed to lower blood lipids, although the result is statistically significant for a certain class of people, the effect is minor. "High-risk men aged 30-69 years should be advised that about 50 patients need to be treated for 5 years to prevent one [cardiovascular] event" (Abramson and Wright, 2007 \cite{abramson2007lipid}). 

For statins side effects and (more or less) hidden risks, see effects in musculoskeletal harm or just pain, Speed et al. (2012) \cite{speed2012statins}. For a general assessment, seeHilton-Jones (2009) \cite{hilton20097}, Hu, Cheung et al. (2012) \cite{hu2012safety}. Roberts (2012) \cite{roberts2012truth} illustrates indirectly various aspects of convexity of benefits, which necessarily implies harm in marginal cases. Fernandez et al. (2011) \cite{fernandez2011statin} shows where clinical trials do not reflect myopathy risks . Blaha et al. (2012) \cite{blaha2012statin} shows "increased risks for healthy patients. Also, Redberg and Katz (2012) \cite{redberg2012healthy}; Hamazaki et al. \cite{hamazaki2012rethinking} : "The absolute effect of statins on all-cause mortality is rather small, if any."
\subsubsection{Other}

For a similar approach to pneumonia, File (2013)\cite{file2013another}.

 Back: Overtreatment  (particularly surgery) for lower back conditions is discussed in McGill (2015) \cite{mcgill2005low}; the iatrogenics (surgery or epidural), Hadler (2009) \cite{hadler2009stabbed}.
 
For a discussion of the application of number needed to treat in evidence-based studies, see Cook et al (1995)  \cite{cook1995number}. One can make the issue more complicated with risk stratification (integrating the convexity to addition of risk factors), see Kannel et al (2000) \cite{kannel2000risk}.  

 Doctor's strikes: There have been a few episodes of hospital strikes, leading to the cancellation of elective surgeries but not emergency-related services. The data are not ample ($n=5$) , but can give us insights if interpreted in \textit{via negativa} manner  as it corroborates the broader case that severity is convex to condition.  It is key that there was no increase in mortality (which is more significant than a statement of decrease). See Cunningham et al. (2008) \cite{cunningham2008doctors} . See also Siegel-Itzkovich (2000) \cite{siegel2000doctors}. On the other hand, Gruber and Kleiner (2010) \cite{gruber2010strikes} show a different effect when nurses strike. Clearly looking at macro data as in Fig. \ref{healthstats} shows the expected concavity: treatment results are concave to dollars invested --life expectancy empirically measured includes the results of iatrogenics.

\section*{Acknowledgment and thanks}
Harry Hong, Raphael Douady, Marco Manca, Matt Dubuque, Jacques Merab, Matthew DiPaola, Christian DiPaola, Yaneer Bar Yam, John Mafi, Michael Sagner, and Alfredo Morales.
%
%

%
%

\end{document}